\documentclass[aps,prd,twocolumn,groupedaddress,nofootinbib,amsmath,amssymb]{revtex4-2}

\usepackage[utf8]{inputenc}

\usepackage{graphicx}
\usepackage[colorlinks=true, allcolors=blue]{hyperref}
\usepackage{physics}
\usepackage{mathtools}
\usepackage{siunitx}
\mathtoolsset{showonlyrefs}

\usepackage{xcolor}
\definecolor{my0}{HTML}{226F99}
\definecolor{my1}{HTML}{F77A3B}
\definecolor{my2}{HTML}{55AD6E}
\definecolor{my3}{HTML}{C795C6}
\definecolor{my4}{HTML}{EBCC36}
\definecolor{mygrey}{HTML}{999999}

\def\[#1\]{\begin{equation}\begin{aligned}[b]#1\end{aligned}\end{equation}} 

\newcommand{\Ogw}{\ensuremath{\Omega_\mathrm{GW}}}

\renewcommand{\H}{\ensuremath\mathcal{H}}
\newcommand{\cssq}{\ensuremath c_\mathrm{s}^2}
\newcommand{\rhocrit}{\ensuremath\rho_\text{cr}}

\renewcommand*{\vec}[1]{\vb{#1}}

\newcommand{\corr}[1]{\Big\langle#1\Big\rangle}
\newcommand{\dirac}[1][3]{\delta^{(#1)}_\mathrm{D}}
\newcommand{\cc}[1]{{#1}^\ast}
\newcommand{\oproj}{\ensuremath \bot}
\newcommand{\pproj}{\ensuremath P}
\newcommand{\Lie}[3][]{\ensuremath\mathcal{L}^{#1}_{#2}{#3}}
\newcommand{\Om}{\ensuremath\Omega_\mathrm{m}}
\newcommand{\Or}{\ensuremath\Omega_\mathrm{r}}

\newcommand{\Ol}{\ensuremath\Omega_\Lambda}

\makeatletter
\newcommand{\trtr}{\@ifstar{\ltrtr}{\utrtr}}
\newcommand{\utrtr}[1]{\ensuremath#1^{\mathrm{TT}}}
\newcommand{\ltrtr}[1]{\ensuremath#1_{\mathrm{TT}}}
\makeatother

\newcommand{\changed}[1]{#1}

\begin{document}
\title{Scalar-Induced Gravitational Waves in a \texorpdfstring{$\Lambda$}{L}CDM Cosmology} 
\author{Marvin Sipp}
\email[]{sipp@thphys.uni-heidelberg.de}
\affiliation{Institut für Theoretische Physik, Universität Heidelberg, Philosophenweg 12, 69120 Heidelberg, Germany}
\affiliation{Astronomisches Rechen-Institut, Zentrum für Astronomie der Universität Heidelberg, Philosophenweg 12, 69120 Heidelberg, Germany}
\author{Björn Malte Schäfer}
\email[]{bjoern.malte.schaefer@uni-heidelberg.de}
\affiliation{Astronomisches Rechen-Institut, Zentrum für Astronomie der Universität Heidelberg, Philosophenweg 12, 69120 Heidelberg, Germany}

\date{17 February 2023}

\makeatletter
\hypersetup{ 
	pdfauthor={Marvin Sipp \& Björn Malte Schäfer},
	pdftitle=\@title
}
\makeatother

\begin{abstract}
We reconsider the gravitational wave spectrum induced by scalar perturbations in spatially flat Friedmann-Lemaître-Robertson-Walker spacetimes, focusing on the matter- and $\Lambda$-dominated epochs. During matter domination, sub-horizon modes are not free and a commonly applied approximation for the derivative of the tensor perturbation is flawed. We show analytically that this leads to a significant overestimation of the energy density spectrum. In addition, we demonstrate that gauge-dependent non-oscillating tensor perturbations appear in the presence of a cosmological constant. Complementing the analytical calculations, we compute the according present-day spectrum numerically for a \textsc{Planck}-like $\Lambda$CDM cosmology, finding that non-oscillating growing modes appear during the transition between matter and $\Lambda$ domination in conformal Newtonian gauge. 
\end{abstract}


\maketitle

\section{Introduction}
The first detection of gravitational waves from binary black hole mergers by the LIGO/Virgo collaborations \cite{LIGO2016a,LIGO2016b} marked the beginning of gravitational wave astronomy. Existing and future detectors like ground-based~\cite{aVIRGO2015,aLIGO2015,Prospects_ALIGO-AVIRGO-KAGRA_2018,ET2020} and space-based~\cite{LISA2017,DECIGO2021} interferometers and pulsar timing arrays~\cite{EPTA2015, NANOGrav2020,IPTA2022} cover a large range of frequencies, including those relevant for stochastic signals of cosmological origin.
The latter could be sourced by, e.\,g., phase transitions, cosmic defects, non-perturbative phenomena like preheating or in the form of primordial fluctuations in inflationary scenarios (see~\cite{Caprini2018} for a review).
An interesting candidate for the detection of a stochastic gravitational wave background is a signal recently reported by the NANOGrav collaboration \cite{NANOGrav2020}\changed{, PPTA \cite{Goncharov2021} and EPTA \cite{Chen2021}}. However, the gravitational wave origin of the signal is yet to be confirmed.
Primordial gravitational waves are particularly interesting, as they could provide decisive evidence for inflation. Their signal can be constrained, e.\,g., by observations of the $B$-mode polarization of the cosmic microwave background (CMB)~\cite{Kamionkowski1997, Seljak1997}. Results from the BICEP2/Keck Array~\cite{BICEP2015} and the \textsc{Planck} Collaboration~\cite{PlanckInflation2018} place upper limits to the tensor-to-scalar ratio, $r\lesssim0.1$, and even tighter constraints when combined.

Regardless of the existence of these more exotic sources, however, \changed{evolving curvature perturbations} generate gravitational waves at second order in perturbation theory (see \cite{Domenech2021_review} for a recent review). This is true for a standard cosmic history, but could also be used to probe e.\,g.\ an early matter-dominated era due to \changed{ultralight} primordial black holes (PBHs)~\cite{Papanikolaou2020,Domenech2021a,Papanikolaou2022}. \changed{Long-lived PBHs are a candidate for Dark Matter (see e.\,g.\ \cite{Carr2016,Carr2020} for reviews) and might form by gravitational collapse during radiation domination if there exist large primordial fluctuations \cite{Zeldovich1967,Hawking1971}. The evolution of the latter would contribute to the induced gravitational wave background and could be used to constrain the PBH abundance \cite{Saito2009,Saito2010}. There is a vast amount of literature on PBHs and their gravitational wave counterpart, and we refer the interested reader to the review articles~\cite{Sasaki2018,Caprini2018,Domenech2021_review} and references therein.}

\changed{For the remainder of this paper, we focus on a standard $\Lambda$CDM cosmology. During late-time matter domination, the first-order evolution of curvature perturbations corresponds to the linear evolution of the density contrast, i.\,e.\ the regime of linear structure formation.} Analytical descriptions of the second-order tensor spectrum were started by \textcite{Ananda2006} and \textcite{Baumann2007} for radiation and matter domination, completed \changed{by \textcite{Espinosa2018,Kohri2018} and semi-analytically extended to transitions between matter and radiation domination by the latter. \textcite{Domenech2020} generalized the analytical computations to general cosmological backgrounds with equation of state parameter ${w\equiv p/\rho \in (0,1]}$}. The tensor perturbations induced by (non-linearly evolving) cosmic large-scale structure can also be obtained from general-relativistic cosmological $N$-body simulations~\cite{Adamek2016a,Adamek2016}.

One interesting finding of the aforementioned studies is a strong enhancement of the gravitational wave spectrum during matter domination. However, the relevant modes do not redshift as one would expect for a radiation fluid. In addition, a strong gauge dependence of these induced tensor perturbations was demonstrated by \textcite{Hwang2017}. \textcite{Domenech2021} could show an approximate gauge independence though, explaining why results for modes sourced during radiation domination agree throughout a large class of gauges. The required assumption that gravitational waves are essentially free on small scales breaks down during matter domination, where they are continuously sourced. \textcite{Ali2021}, however, could show that computations agree in seven different gauges when only the oscillating part of the tensor perturbation is considered. As an aside, they mention that no non-oscillating contribution exists in conformal Newtonian gauge when working with the proper definition of the gravitational wave energy density. Indeed, in large parts of previous literature the latter is defined in a way implicitly assuming that gravitational waves are free. Since this approximation is flawed during matter domination, however, we reconsider the scalar-induced gravitational wave spectrum in cosmologies dominated by matter and a cosmological constant in this paper. While we focus on a standard \textsc{Planck}-like $\Lambda$CDM model, our analytical solutions can also be applied to non-standard scenarios like an early matter domination.

We recapitulate the formalism for scalar-induced gravitational waves in Section~\ref{sec:formalism}, including a brief discussion of the gauge issue in Section~\ref{sec:gauge}. Thereafter, analytical solutions for a matter dominated epoch are presented in Section~\ref{sec:matter_dom}, followed by the (approximately) de~Sitter case in Section~\ref{sec:de_dom}. In Section~\ref{sec:numerical_results}, we present numerical solutions for a standard cosmological model with a spatially flat background and a cosmological constant in order to correctly account for transitions between the cosmological eras.

\section{Induced Gravitational Waves}\label{sec:formalism}
The background spacetime of the cosmological standard model ($\Lambda$CDM) is of the well-known Friedmann-Lemaître-Roberson-Walker (FLRW) type with a cosmological constant $\Lambda$, populated by cold Dark Matter (CDM) and a small fraction of baryons~\cite{Bartelmann2010}. It is constructed under the assumptions of spatial homogeneity and isotropy, well justified by large-scale observations of the cosmos, e.\,g.\ CMB measurements~\cite{Planck2018Overview}. Einstein's field equations then simplify to Friedmann's equations, which describe the dynamics of spacetime in terms of a scale factor $a$. 
Under these assumptions, the cosmic fluid energy-momentum tensor, as seen from a comoving observer, is that of a perfect fluid. Here, we will restrict ourselves to a spatially flat background.

Evidently, the Universe does not exactly obey the aforementioned Robertson--Walker symmetries, as attested by the existence of cosmic structures. The latter already appear at first order in perturbation theory, where \emph{scalar}, divergence-free \emph{vector} and traceless and transverse \emph{tensor} modes evolve independently~\cite{Bardeen1980,Stewart1990}. At linear order, vector modes decay quickly and tensor perturbations are sourced by anisotropic stress, which vanishes necessarily for a perfect cosmic fluid~\cite[see e.\,g.][for excellent reviews]{Mukhanov1992,Malik2009}. Therefore, only first-order scalar perturbations are considered in our study. They can be parameterized by the gauge-invariant Bardeen potential $\Phi$ as a generalization of the Newtonian gravitational potential~\cite{Bardeen1980}.
At higher order in perturbation theory, mode mixing can occur, transferring amplitudes of excitations between scalar and tensorial perturbations. In order to investigate gravitational waves induced by linearly evolving cosmic \changed{perturbations}, one thus has to go beyond linear order.

In conformal Newtonian gauge, the second-order metric takes the form
\[\label{eq:second_order_metric}
	g = a^2\qty{-\qty(1+2\Phi) \dd{\eta}^2 + \qty[\qty(1-2\Phi)\gamma_{ij}+\frac{1}{2}h_{ij}] \dd x^i \dd x^j},
\]
where $\gamma_{ij}$ denotes the Euclidean metric and all second-order modes but the traceless-transverse tensor $h_{ij}$ were discarded.\footnote{Throughout this work we denote spatial and spacetime indices with Latin and Greek letters, respectively. We also use Einstein's summation convention. Furthermore, we work in units where the speed of light equals unity, $c=1$.} This is justified by the fact that $n$-th order scalar, vector and tensor perturbations still decouple, only mixing with lower-order variables~\cite{Malik2009}. Starting instead with first-order tensor perturbations and second-order scalar modes, for example, one could investigate signatures of primordial gravitational waves on the density contrast of cosmic large-scale structure~\cite{Tomita1971,Matarrese1998,Bari2022}.

Inserting the above metric into Einstein's field equations, together with the perfect fluid energy-momentum tensor, they yield the second-order gravitational wave equation~\cite{Ananda2006,Baumann2007,Acquaviva2003},
\[\label{eq:second_order_gws}
	h_{ij}''+2\H h_{ij}'-\laplacian h_{ij} = 4\,\trtr{\mathcal{S}_{ij}}.
\]
On the right-hand side, the superscript $\trtr{}$ denotes taking the traceless and transverse part and $\H\equiv a'/a$ is the conformal Hubble rate, with the prime indicating derivatives with respect to conformal time $\eta$. The source term is given by
\[\label{eq:source}
	\mathcal{S}_{ij} = \kappa a^2\qty(\bar\rho+\bar{p})\, \delta u_i \delta u_j - 4\,\Phi\,\partial_i\partial_j\Phi - 2\,\partial_i\Phi\,\partial_j\Phi,
\]
where $\kappa = 8\pi G$ and $\bar{\rho}$, $\bar{p}$ are the background density and pressure, respectively. $\delta u^i$ is the first-order velocity perturbation of the fluid. The second and third term of~\eqref{eq:source} are effective sources arising in the perturbative expansion of the Einstein tensor. This is precisely how first-order perturbations induce higher-order modes.

The first-order perturbation variables evolve linearly, and it is useful to separate the Bardeen potential into an initial perturbation and a transfer function, ${\Phi(\vec{k},\eta) = T_\Phi(k,\eta)\, \phi_{\vec{k}}}$. Furthermore, $\delta\vec{u} \parallel \vec{k}$ under the aforementioned assumptions (cf.\ Appendix~\ref{app:first_order_pert}), such that we can write ${\delta\vec{u}(\vec{k},\eta) = T_u(k,\eta)\,\phi_{\vec{k}}}\,\vec{k}$. In Fourier space, the source reads
\begin{multline}\label{eq:source_fourier_short}
	\trtr{\mathcal{S}}_{ij}(\vec{k}, \eta) = \oproj_{ij}{}^{ab}(\hat{\vec{k}})\\\times\int_{\mathbb{R}^3}\frac{\dd[3]q}{(2\pi)^3}\, q_a\,q_b\,\phi_{\vec{q}}\,\phi_{\vec{k}-\vec{q}}\,f(q,\abs{\vec{k} - \vec{q}},\eta),
\end{multline}
with the expression
\begin{multline}\label{eq:def_f}
	f(q,p,\eta) \equiv{} -3(1+w)\mathcal{H}^2\,T_u(q,\eta)\,T_u(p, \eta)\\ + 2\,T_\Phi(q,\eta)\,T_\Phi(p, \eta),
\end{multline}
where $w(\eta) \equiv \bar p(\eta)/\bar\rho(\eta)$ denotes the background equation of state parameter.
The definition of the traceless-transverse projector is adopted from \textcite{Caprini2018},
\[
	\oproj_{ijlm}(\hat{\vec{k}}) &\equiv \pproj_{il} \pproj_{jm} - \frac{1}{2} \pproj_{ij} \pproj_{lm} \\\text{with}\quad \pproj_{ij}(\hat{\vec{k}}) &\equiv \delta_{ij} - \hat{k}_i \hat{k}_j,
\]
where a hat indicates unit vectors. 
At first order in relativistic perturbation theory, the potential and velocity transfer functions can be related (via~\eqref{eq:rel_vel} in Appendix~\ref{app:first_order_pert}),
\[\label{eq:Tu_rel}
	T_u(k,\eta) = \frac{2i}{3(1+w)\mathcal{H}^2} \qty(T'_\Phi+\H T_\Phi).
\]
Equation~\eqref{eq:def_f} can thus be written as~\cite{Baumann2007}
\[\label{eq:f_rel}
	f(q, p, \eta)
	={}& \qty(\frac{4}{3(1+w)}+2) T_\Phi(q,\eta)T_\Phi(p, \eta)\\
	&+ \frac{4}{3(1+w)\H^2} T'_\Phi(q,\eta)T'_\Phi(p, \eta)\\
	&+ \frac{4}{3(1+w)\H} \Big(T_\Phi(q,\eta)T'_\Phi(p, \eta)\\
	&+ T_\Phi(q,\eta)T'_\Phi(p, \eta)\Big).
\]
In Fourier space,~\eqref{eq:second_order_gws} can be solved using a suitable Green's function $g_k(\eta,\bar\eta)$, such that (see Appendix~\ref{app:greens})

\[\label{eq:h_conv}
	h_{ij}(\vec{k}, \eta) = \frac{4}{a(\eta)} \int_{\eta_\mathrm{i}}^\eta \dd\tilde{\eta}\  g_k(\eta,\tilde{\eta})\,a(\tilde\eta)\,\trtr{\mathcal{S}}_{ij}(\vec{k},\tilde\eta),
\]
where $\eta_\mathrm{i}$ is some time where initial conditions are specified.

\subsection{Gravitational Wave Spectra}
In the context of cosmological scalar perturbations, the sources are random fields and thus characterized by their statistics. Therefore, a similar treatment is indicated for the induced stochastic background of gravitational waves.
As usual, a power spectrum can be defined in terms of the Fourier-transformed two-point correlator,
\[
	\corr{h_{ij}(\vec{k}, \eta)\ {\cc{h}}^{ij}(\tilde{\vec{k}},\eta)} \equiv (2\pi)^3\ P_h(k, \eta)\ \dirac(\vec{k}-\tilde{\vec{k}}),
\]
where $\dirac$ denotes the Dirac distribution.
Substituting the expressions above and using Wick's theorem, one finds~\cite[e.\,g.][]{Ananda2006,Baumann2007,Kohri2018}
\begin{multline} \label{eq:P_h_full}
	P_h(k, \eta) ={} \frac{4}{\pi^2} \int_{0}^\infty \dd q \int_{-1}^{+1} \dd\mu\ q^6\qty(1-\mu^2)^2\,\\\times\mathcal{I}^2(k,q,\mu,\eta)\,P_\phi(q)\,P_\phi\qty(\abs{\vec{k-\vec{q}}}),
\end{multline}
with the initial scalar power spectrum $P_\phi$ and the kernel
\[\label{eq:kernel}
	\mathcal{I}(k,q,\mu,\eta) \equiv \int_{\eta_\mathrm{i}}^\eta \dd\tilde\eta\  g_k(\eta,\tilde\eta)\,\frac{a(\tilde\eta)}{a(\eta)}\,f\qty(q, \abs{\vec{k}-\vec{q}}, \tilde\eta).
\]

If one were to go to higher order in perturbation theory, an effective energy-momentum source would be contributed by the gravitational waves themselves. Even though this energy cannot be localized in General Relativity, an effective energy density can be defined by averaging over a suitably large volume. Sufficiently below the horizon, it reads~\cite{Maggiore2000}
\[\label{eq:rho_GW}
	\rho_\text{GW}(\vec x, \eta) = \frac{\corr{h'_{ij}(\vec x, \eta) h'^{ij}(\vec x, \eta)}}{32\pi G a^2},
\]
taking the typical form of a kinetic term, i.\,e. the energy (density) of a massless excitation.
Equivalently, using the conventions of e.\,g.\ \textcite{Watanabe2006}, the dimensionless energy density spectrum reads
\[\label{eq:def_Omega}
	\Omega_\text{GW}(k, \eta) \equiv \frac{1}{\rhocrit}\dv{\rho_\text{GW}}{\log k},
\]
with the cosmological critical density ${\rhocrit = 3\,\H^2/(\kappa a^2)}$. In our model, this yields\footnote{The temporal dependence of $h_{ij}$ is completely absorbed in the kernel $\mathcal{I}$.} 
\begin{multline}\label{eq:Omega}
	\Omega_\text{GW}(k, \eta) = \frac{4\,k^3}{3\pi\H^2(\eta)} \int_{0}^\infty \dd q \int_{-1}^{+1} \dd\mu\ q^6\qty(1-\mu^2)^2\\ \times {\mathcal{I}'}^2(k,q,\mu,\eta)\,P_\phi(q)\, P_\phi\qty(\abs{\vec{k}-\vec{q}}).
\end{multline}
This expression can now be evaluated for any given initial power spectrum of scalar perturbations and their temporal evolution.
It shall be noted that in literature gravitational waves are commonly assumed to be freely propagating, i.\,e.\ without significant sourcing or resonance effects. In this case, one has $h_{ij}' \propto k\,h_{ij}$ and thus
\[\label{eq:Omega_approx}
	\Omega_\mathrm{GW}(k,\eta)\changed{\bigg\vert_\mathrm{free}} \simeq \frac{\pi}{3}\frac{k^5\,P_{h}(k,\eta)}{\H^2(\eta)}.
\]

In a cosmology dominated by matter or Dark Energy, however, this assumption is invalid and the full expression~\eqref{eq:Omega} has to be used instead. We will quantify the importance of this analytically in Section~\ref{sec:analytical_results} and numerically in Section~\ref{sec:numerical_results}. Before, the issue of gauge ambiguity shall briefly be addressed.

\subsection{The Gauge Issue}\label{sec:gauge}
The metric~\eqref{eq:second_order_metric} is valid in the conformal Newtonian gauge. This choice, however, is arbitrary. In fact, the group of gauge transformations is isomorphic to the diffeomorphism group in General Relativity~\cite{Trautman1970,Trautman1979}. It is therefore directly related to the the freedom of coordinate choice, which is paramount to the theory. Any one-parameter family of diffeomorphisms can be expressed as a (generally infinite-rank) one-parameter family of \emph{knight diffeomorphisms}~\cite{Bruni1997}. The latter are compositions of one-parameter \emph{groups} of diffeomorphisms that are each the flow of a generating vector field. This construction allows a perturbative expansion of coordinate transformations and thus a well-defined notion of $n$-th order gauge transformations.

Under a gauge transformation $\phi_\epsilon$ that is parameterized by an expansion parameter $\epsilon$, any tensorial quantity $\mathcal{Q}$ transforms as~\cite{Bruni1997}
\[
	\tilde{\mathcal{Q}} &\equiv (\phi_\epsilon)^\ast \mathcal{Q}\\
	&= \mathcal{Q} + \epsilon\Lie{X^{(1)}}{\mathcal{Q}} + \frac{\epsilon^2}{2}\qty(\Lie[2]{X^{(1)}}{} + \Lie{X^{(2)}}{})\mathcal{Q} + \dots,
\]
where an asterisk denotes the pullback and $\Lie{}{}$ the Lie derivative. The generators $X^{(i)}$ are four-vector fields and thus each have two scalar and one vector, but no tensorial degree of freedom.
One can therefore immediately see that first-order gravitational waves are gauge-invariant. As discussed above, however, scalar and vector degrees of freedom enter the second-order (and higher) gravitational wave equation, making the gauge issue more intricate: 
While the tensor perturbation is not directly affected by the transformation, the sources are. Therefore, only free\footnote{Here, \emph{free} means that both real and effective sources vanish.} gravitational waves are gauge-invariant at higher order. This is at the heart of the gauge issue of scalar-induced gravitational waves, which was made explicit by \textcite{Hwang2017}, who found vast differences in their power spectra in different gauges.

One can, of course, construct a gauge-invariant second-order tensor variable. As \textcite{Domenech2021} point out, however, there are infinitely many ways of doing so, and there is no \emph{a priori} preferred choice and it remains unclear how to connect the constructed variable to physical observables. There are different ways of approaching this ambiguity:
\textcite{DeLuca2020}, for example, try to choose a preferred gauge from the perspective of a gravitational wave \changed{laser interferometer, arguing} in favor of synchronous gauge. This choice, however, suffers from a relevant residual gauge freedom \cite{Lu2020}. As an alternative approach, \textcite{Domenech2021} define a class of \emph{reasonable gauges} in which a detector should not oscillate in absence of physical gravitational waves. This includes conformal Newtonian gauge that we use here and emphasises the correspondence between the detector and a fundamental observer in a FLRW universe. They show that within this class, the gravitational wave energy density spectrum is gauge-invariant and decays as expected for a radiation fluid, provided that the sources are practically inactive. On sub-horizon scales, this is often the case, as e.\,g.\ in a radiation-dominated universe the sourcing of a mode is strongest at horizon crossing and quickly decays afterwards~\cite{Baumann2007}. This approximate gauge-invariance can be well understood in light of the aforementioned gauge-invariance of free gravitational waves: In a reasonable class of gauges, the effective source terms vanish whenever the physical sources do. The gravitational wave equation then becomes purely tensorial and is thus unaffected by any knight diffeomorphism, i.\,e.\ gauge transformation.

There is, however, one caveat to the above argument: In cosmological epochs where $\cssq = 0$, e.\,g.\ matter or Dark Energy domination, the sources become constant in terms of the Bardeen potentials and therefore remain active. Indeed, this leads to the appearance of gauge artifacts and a non-oscillatory contribution to the tensor perturbations. Arguing that only oscillating tensor modes can be regarded as \emph{gravitational waves}, \textcite{Inomata2020} show that the kernel $\mathcal{I}$ (and thus $\Omega_\mathrm{GW}$) coincides up to non-oscillatory terms during matter domination in conformal Newtonian and synchronous gauge, and \textcite{Ali2021} extend this result to seven different gauges.
Such a distinction between oscillating and non-oscillating contributions is reasonable, as at second order in perturbation theory gauge transformations simply add terms to the kernel~\cite{Lu2020}. For the well-behaved class of gauges mentioned above, the gauge artifacts should be non-oscillating. Both strain and energy density of the oscillating gravitational wave part of the tensor perturbation are then gauge-invariant within this class.

In addition, the authors of~\cite{Ali2021} mention that the only non-oscillatory term appearing in conformal Newtonian gauge is constant and thus automatically disappears if one uses~\eqref{eq:Omega} instead of~\eqref{eq:Omega_approx} when computing $\Omega_\mathrm{GW}$. We want to stress, however, that this is not merely a useful trick, but that using~\eqref{eq:Omega_approx} implicitly employs the approximation $h_{ij}' \propto k\,h_{ij}$, which is invalid as long as sources are active---only free gravitational waves obey this relation. Nonetheless, this form of $\Omega_\mathrm{GW}$ is widespread in literature, even when matter-dominated universes are discussed. With this in mind, we want to reconsider the scalar-induced gravitational wave spectrum in different cosmological epochs in the following sections.

\section{Analytical Tensor Spectra}\label{sec:analytical_results}
For a constant equation of state of the cosmic fluid and adiabatic perturbations, analytical solutions for the gravitational wave spectrum can be found.
First analytical investigations were made for radiation and matter domination~\cite{Ananda2006,Baumann2007,Assadullahi2009,Espinosa2018}. Transitions between these epochs were modeled in a semi-analytical way by \textcite{Kohri2018}, and it was shortly after shown that the spectrum also depends on the specifics of the transition~\cite{Inomata2019,Inomata2019a}.
More recently, a general solution for $0 < w \leq 1$ was given by \textcite{Domenech2020}.

During radiation domination, the scalar source $\Phi$ is suppressed as $(k\eta)^{-2}$, so the approximation of free sub-horizon gravitational wave modes is valid. Therefore, we will not focus on this case in more detail and refer to previous literature, where it was found that for a flat primordial curvature power spectrum, the induced gravitational wave spectrum is also of constant amplitude~\cite{Ananda2006,Baumann2007,Kohri2018}. In a standard $\Lambda$CDM cosmology, we therefore expect the ``UV tail'' emitted during radiation domination to be flat. As discussed above, however, gravitational waves are not free on any scale when the scalar perturbations are dominated by matter with $\cssq = 0$.

\subsection{Matter Domination}\label{sec:matter_dom}
During (pressureless) matter domination, both the cosmic equation of state and the speed of sound vanish, $w=\cssq=0$.
The first-order equation~\eqref{eq:scalar_evolution} for the evolution of scalar perturbations then simplifies to
\[
	\Phi'' + \frac{6}{\eta} \Phi' = 0,
\]
which has an irrelevant decaying and a constant solution. Normalizing $T_\Phi$ to unity during matter domination,~\eqref{eq:f_rel} simply reduces to
\[\label{eq:f_matter}
	f(q,p,\eta) = \frac{10}{3}.
\]
Sufficiently below the horizon and during the matter-dominated era, a Newtonian treatment of cosmic structure should be applicable (see Appendix~\ref{app:first_order_pert}).
However, one should evolve the primordial \changed{curvature} power spectrum to the present day using the fully relativistic transfer functions and only employ the Newtonian model to work backwards from there.
Otherwise, relevant processes in the radiation-dominated Universe are neglected and the observed present-day matter power spectrum is not recovered.
Indeed, combining the linearized continuity equation~\eqref{eq:lcontinuity} and the Poisson equation~\eqref{eq:poisson}, assuming a purely matter-dominated background density $\bar{\rho}=\Om\rhocrit a^{-3}$, one obtains the relation
\[
	T_u = \frac{2ai}{3\,\Om H_0^2} \qty(T_\Phi'+\H T_\Phi).
\]
During matter domination, where $w=0$ and ${\H^2 = H_0^2\,\Om a^{-1}}$, this exactly coincides with the relativistic relation~\eqref{eq:Tu_rel}. Furthermore, the potential transfer function is related to the density perturbation via the Poisson equation. With the same normalization as before, it can be expressed in terms of the growth function $D_+$ as $T_\Phi = D_+/a$, which is unity during matter domination. Thus,~\eqref{eq:f_matter} is recovered in the Newtonian approximation. While expected on small scales, this shows that the Newtonian result coincides with the relativistic one which was derived without the aforementioned restriction.

\begin{figure}
	\includegraphics[width=\linewidth]{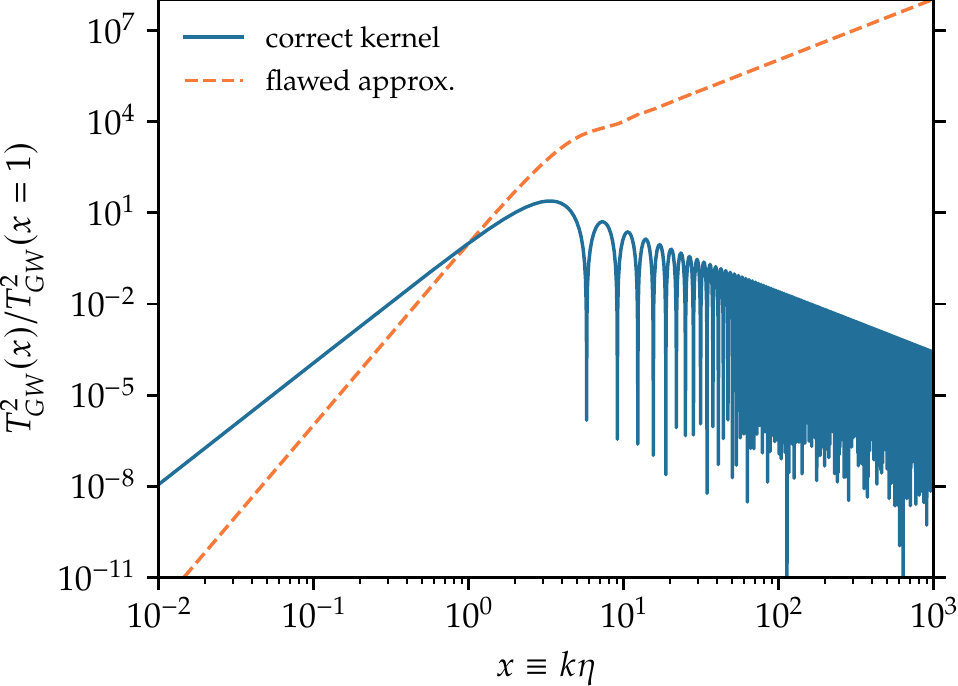}
	\caption{Comparison of the correct time evolution of $\Omega_\mathrm{GW}$ during matter domination (\textcolor{my0}{\bfseries blue}, solid) with the incorrectly approximated result (\textcolor{my1}{\bfseries orange}, dashed) commonly used in literature, both normalized to unity at horizon scale, $x = 1$. The correct solution decays as $\eta^{-2}\propto a^{-1}$ on sub-horizon scales, as expected for gravitational waves.}\label{fig:Omega_MD_comparison}\label{fig:Omega_MD_evol}
\end{figure}

With the appropriate Green's function \eqref{eq:greens_mat_lam}, the kernel~\eqref{eq:kernel} can be integrated analytically, yielding~\cite[cf.][]{Kohri2018}
\[\label{eq:kernel_MD}
	\mathcal{I}(k,\eta) = \frac{10}{3}\qty(\frac{1}{k^2} + 3\,\frac{k\eta\cos (k\eta)-\sin(k\eta)}{k^5\eta^3}).
\]
Splitting $\Omega_\mathrm{GW}$ into a purely spatial and a time-dependent part, ${\Omega_\mathrm{GW}(k,\eta) = T^2_\mathrm{GW}(k,\eta)\,\omega_\mathrm{GW}(k)}$, we arrive at
\[\label{eq:Omega_MD_true}
	T^2_\mathrm{GW}(k,\eta) &\equiv \qty(\frac{\mathcal{I}'(k,\eta)}{\H(\eta)})^2 \\
	&=  \frac{25\,\qty(3\,k\eta\cos(k\eta)+(k^2\eta^2-3)\sin(k\eta))^2}{k^{10}\eta^6}.
\]
The leading order in a Taylor expansion, which gives the superhorizon behavior, is proportional to $\eta^4 \propto a^2$, while the dominant term ignoring oscillations is of order $\eta^{-2} \propto a^{-1}$. 
This is consistent with the fact that the sourcing is peaked around horizon entry and afterwards the gravitational waves, being massless excitations following null geodesics, should redshift like any radiation fluid. Interestingly, this contradicts the results first obtained by \textcite{Baumann2007}, where it is claimed that second-order gravitational waves induced during matter domination do \emph{not} redshift. Their results are recovered, however, when employing~\eqref{eq:Omega_approx}, which gives
\[\label{eq:kernel_MD_false}
	T^2_\mathrm{GW}(k,\eta)\changed{\bigg\vert_\mathrm{free}} &\equiv \qty(\frac{k\mathcal{I}(k,\eta)}{\H(\eta)})^2 \\
	&= \frac{25\,\qty(3\,k\eta\cos(k\eta) - 3\,\sin(k\eta)+k^3\eta^3)^2}{9\,k^{8}\eta^4}.
\]
Consequently, the first order in Taylor expansion would be proportional to $\eta^6 \propto a^3$ and the late-time behavior to $\eta^2 \propto a$. A comparison of the two results, both normalized to unity at $k\eta = 1$, is shown in Figure~\ref{fig:Omega_MD_evol}.

\textcite{Ali2021}, who define gravitational waves as the oscillatory part of the tensor perturbation only, note that the difference is due to a non-oscillatory contribution---the $k^{-2}$-term in~\eqref{eq:kernel_MD}---which automatically vanishes in conformal Newtonian gauge when applying the time derivative. For gauges where the non-oscillatory term is time-dependent, the authors explicitly discard it. Alternatively, the non-oscillatory contribution can be removed by a gauge transformation~\cite{Domenech2021}.

We want to stress, however, that this contribution is \emph{not} simply a gauge mode, but a consequence of an invalid approximation: As mentioned earlier, employing ${\mathcal{I}' \propto k\,\mathcal{I}}$ is only valid for free waves. During matter domination, however, the sources remain constant, rendering the approximation inapplicable. This is important, as non-oscillatory gauge-dependent contributions to the tensor perturbation are not necessarily irrelevant: When doing calculations in some arbitrary gauge, they cannot simply be discarded, even if one \emph{defines} gravitational waves as purely oscillatory. What is actually relevant is the response of some detector. Employing the aforementioned approximation outside its scope of validity will lead to erroneous contributions to the second-order tensor perturbations. While ubiquitous in literature, it should therefore be used with caution and the physical approximation in mind, resorting to the full expression (i.\,e.~\eqref{eq:Omega} in our case) when necessary.

\changed{The above discussion is also an argument in favor of conformal Newtonian gauge for the computation of \Ogw, as in that case only oscillating modes (which typical detectors are sensitive to) contribute to the energy density up until matter domination. Any gauge transformation would transform the kernel as~\cite{Lu2020,Ali2021}
\[
	\mathcal{I}(k,q,\mu,\eta) \to \mathcal{I}(k,q,\mu,\eta) + \delta\mathcal{I}(k,q,\mu,\eta),
\]
where $\delta\mathcal{I}(k,q,\mu,\eta)$ is in general time-dependent and thus does not vanish upon applying a time derivative. Furthermore, as discussed in Section~\ref{sec:gauge}, the additional terms will be non-oscillatory for a large class of gauges. Transforming from conformal Newtonian to synchronous gauge, for example, we would have\footnote{Note that our convention for $\mathcal{I}$ differs by a factor of $25/9\ k^{-2}$ from \cite{Ali2021}.}~\cite{Ali2021}
\[
	\delta\mathcal{I}(k,q,\mu,\eta) = \frac{\eta^2}{72}\qty[\qty(q^2-kq\mu)\,\eta^2-44].
\]
The kernels for matter domination in seven different gauges can be found in Ref.~\cite{Ali2021}.
}

So far, the discussion has been independent from the initial matter power spectrum. Consider now a flat power spectrum $k^3\,P_\phi(k) = \text{const}$. The spatial integrals in~\eqref{eq:P_h_full} and~\eqref{eq:Omega} are divergent in this case. Furthermore, very small wavelengths are in the non-linear regime of structure formation and the second-order perturbative approach would become insufficient. We will therefore employ a UV cutoff $k_\mathrm{UV}$~\cite{Assadullahi2009,Kohri2018},
\[
	\frac{k^3}{2\pi^2}\,P_\phi(k) = \frac{4}{9}\,\Delta_\mathcal{R}^2(k_0)\,\Theta(k_\mathrm{UV}-k),
\]
with the Heaviside function $\Theta$ and a dimensionless normalization $\Delta_\mathcal{R}^2(k_0)$. In practice, matter domination is not past-eternal anyway, and the cutoff scale arises naturally during the smooth transition between radiation and matter domination. In order to keep the results of this section more general, we will not specify an explicit value for $k_\mathrm{UV}$ here.

\begin{figure}
	\includegraphics[width=\linewidth]{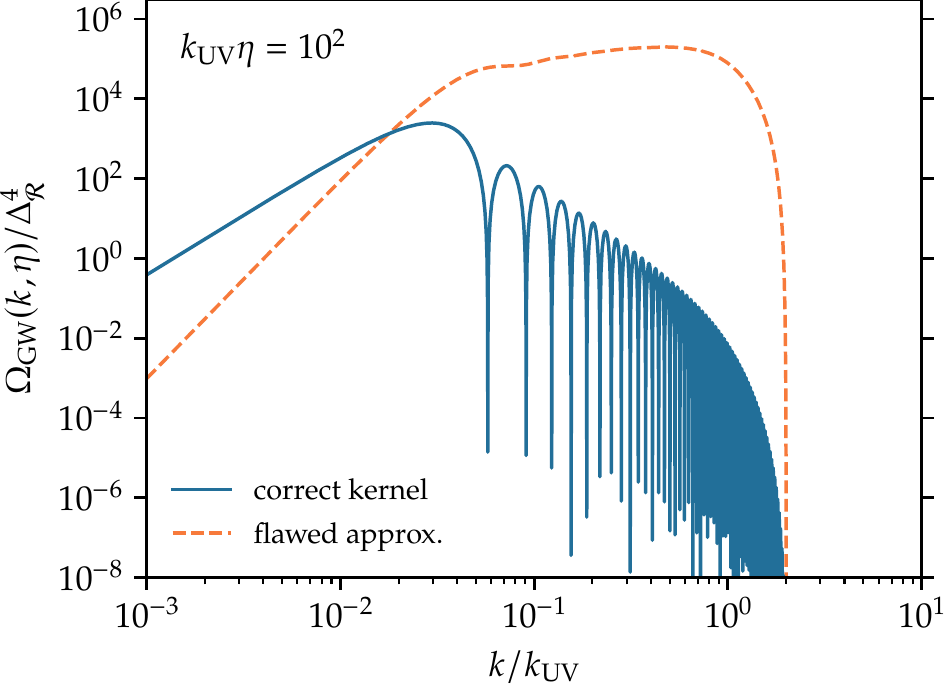}
	\caption{Analytical solution for the gravitational wave energy density spectrum $\Omega_\mathrm{GW}(k,\eta)$ induced in an (ongoing) matter-dominated era assuming a scale-invariant primordial matter power spectrum. A sharp UV cut-off $k_\mathrm{UV}$ was imposed and the time set to $k_\mathrm{UV}\eta = 10^2$. Beyond the cutoff scale, a realistic spectrum would transition to the flat spectrum produced during a preceding radiation-dominated era. For comparison, both the improved result (\textcolor{my0}{\bfseries blue}, solid) and the one using a commonly applied though flawed approximation for the kernel is shown (\textcolor{my1}{\bfseries orange}, dashed).}\label{fig:Omega_MD}
\end{figure}

Following the calculation of \textcite{Kohri2018}, we arrive at
\begin{widetext}
	\[\label{eq:omega_spatial_cutoff}
		\omega_\mathrm{GW}(k) = \Delta_\mathcal{R}^4\,k^4\frac{16\pi^3}{25515}
		\begin{cases}
			105\,\tilde k^2 + 768\,\tilde{k} - 2520 + 1792\,\tilde{k}^{-1} & (0<\tilde{k}\leq1)\\
			105\,\tilde k^2 - 768\,\tilde{k} + 1960 - 1792\,\tilde{k}^{-1} + 896\,\tilde k^{-4} - 256\,\tilde{k}^{-6} & (1<\tilde{k}\leq2)\\
			0 & (2 < \tilde{k} < \infty)
		\end{cases},
	\]
\end{widetext}
where $\tilde{k} \equiv k/k_\mathrm{UV}$. The complete energy density spectrum,
\[
	\Omega_\mathrm{GW}(k,\eta) = T^2_\mathrm{GW}(k,\eta)\,\omega_\mathrm{GW}(k),
\]
is shown in Figure~\ref{fig:Omega_MD} at a time $k_\mathrm{UV}\eta = 10^2$ sufficiently after the onset of matter domination, for both the correctly evaluated time-dependent part~\eqref{eq:kernel_MD} and the flawed approximation~\eqref{eq:kernel_MD_false}. The latter case is the result commonly encountered in literature~\cite{Kohri2018}.
Clearly, the erroneous non-oscillating contributions introduced by the flawed approximation of the kernel strongly dominate over the approximately gauge-invariant oscillating part, leading to a strong overestimation of the spectrum. It is particularly pronounced close to $k_\mathrm{UV}$, which is in accordance with the asymptotic behavior mentioned before. Qualitatively, the spectrum computed with the exact definition of $\Omega_\mathrm{GW}$ reaches its maximum near horizon entry and redshifts thereafter, while the invalidly approximated result grows towards the cutoff. This explains the strong maximum at the scale of matter-radiation equality found by \textcite{Baumann2007}.

\changed{Albeit we focus on a standard $\Lambda$CDM model in this work, the analytical considerations above do not depend on this assumption and equally apply to an early matter-dominated era. However, as pointed out in Ref.~\cite{Domenech2021}, the tensor perturbations generated during early matter domination \emph{do} propagate freely in a subsequent era after the sources have decayed, and initially non-oscillating contributions could start to oscillate. Nevertheless, the observable signal has to be gauge-independent, so differences between gauges should be eliminated in the transition from early matter domination to the subsequent era. Indeed, it has been shown that the specifics of this transition may significantly impact the spectrum \cite{Inomata2019,Inomata2019a}. A more thorough investigation of an early matter-dominated scenario in light of the flawed free-wave approximation is thus left for subsequent research.}

\subsection{Dark Energy Domination}\label{sec:de_dom}
A de Sitter universe is perfectly homogeneous, as the cosmological constant does not cluster. In this section, however, we will consider a universe whose background evolution is driven by the cosmological constant~(${w=-1}$), while matter fluctuations with $\cssq = 0$ are present. The first term of~\eqref{eq:def_f} vanishes and only effective source terms remain. Again, the same is obtained in Newtonian cosmology, as the linear growth factor obeys $D_+'=0$ in this case. With $\H \propto a$, i.\,e.\ $a\propto\eta^{-1}$ and the dominant behavior of the potentials, $\Phi\propto a^{-1}$, we have $f(\eta)\propto\eta^2\propto a^{-2}$. For the kernel, we obtain
\[
	\mathcal{I}(k,\eta) \propto \frac{\eta^2}{k^2} + 2\,\frac{1-k\eta\sin (k\eta)-\cos(k\eta)}{k^4},
\]
such that the temporal evolution of the energy density power spectrum reads
\[
	T^2_\mathrm{GW}(k,\eta) \equiv \qty(\frac{\mathcal{I}'(k,\eta)}{\H(\eta)})^2 \propto \frac{4\,\eta^4}{k^4}\sin^4\mathopen{}\qty(\frac{k\eta}{2})\mathclose{}.
\]
The wavelength of oscillations is increased by a factor of two compared to the gravitational wave expectation, which can be traced back to the non-oscillating term in the kernel. In contrast to the matter-dominated case it does not drop out, as it is time-dependent \changed{even in conformal Newtonian gauge}. Indeed, following the reasoning of~\cite{Ali2021} and the physical arguments as to why a this particular term should not contribute, we instead obtain
\[
	T^2_\mathrm{GW}(k,\eta) \propto \frac{4\,\eta^4\,\cos[2](k\eta)}{k^4}.
\]
Physically, the critical density freezes during cosmological constant domination, such that the energy density of a radiation fluid should decay proportional to $a^{-4}$, which the above result fulfills. In addition, the gravitational wave frequency redshifts until the mode ceases to oscillate upon crossing the shrinking comoving horizon. For a flat primordial matter power spectrum, one could introduce a UV cutoff similarly to the above treatment of matter domination, but the spatial integration and thus~\eqref{eq:omega_spatial_cutoff} would remain unchanged.

In order to compute the late-time induced gravitational wave spectrum in a $\Lambda$CDM universe, one would now need to investigate the transition between matter and cosmological constant domination. A semi-analytical approach similar to the one in~\cite{Kohri2018}, where the transition between matter and radiation domination was investigated, is more complicated here due to the transition from a growing to a decaying relationship between conformal time and redshift. We shall not pursue this further here and leave it for future research. The details of such a transition, however, may have a significant impact on the spectrum~\cite{Inomata2019,Inomata2019a}.

\section{Numerical Results for a Planck-like \texorpdfstring{$\Lambda$}{L}CDM cosmology}\label{sec:numerical_results}
\begin{figure}
	\includegraphics[width=\linewidth]{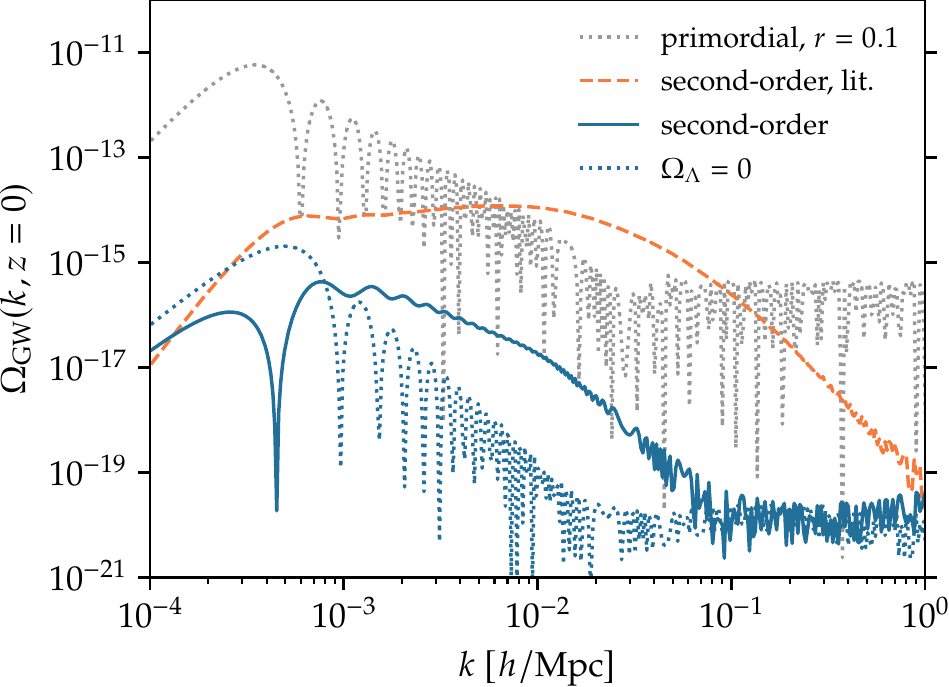}
	\caption{Present-day tensor spectrum $\Omega_\mathrm{GW}(k, z=0)$ for a \textsc{Planck}-like cosmology (\textcolor{my0}{\bfseries blue}, solid), similarly for $\Omega_\Lambda = 0$ (\textcolor{my0}{\bfseries blue}, dotted) and the results obtained using the approximation $h_{ij} \propto k\,h_{ij}$ that is not justified during matter domination (\textcolor{my1}{\bfseries orange}, dashed). For comparison, the primordial first-order spectrum for a simple inflationary model with optimistic tensor-to-scalar-ratio $r=0.1$ is shown (\textcolor{mygrey}{\bfseries gray}, dotted)~\cite[cf.][]{Watanabe2006}.}\label{fig:Omega_z0}
\end{figure}

In order to account for the transition between the cosmological epochs in a realistic model, we integrate~\eqref{eq:Omega} numerically for a $\Lambda$CDM cosmology with \textsc{Planck}-like parameters and a slightly red-tilted initial matter power spectrum~\cite{PlanckResults2018}. The potential transfer function is obtained from the numerical Boltzmann solver CLASS \cite{Class2011}. In Figure~\ref{fig:Omega_z0}, the present-day tensor energy density spectrum $\Omega(k,z=0)$ is shown, both using the full definition~\eqref{eq:Omega} and the approximated expression~\eqref{eq:Omega_approx}. In the latter case, our results are compatible with Ref.~\cite{Baumann2007}. As expected from the analytical discussion of Section~\ref{sec:matter_dom}, however, the power on large and intermediate scales is strongly overestimated by the invalid free-wave assumption $h_{ij}'\propto k\,h_{ij}$. Indeed, while previous calculations find that the second-order signal dominates over an optimistic\footnote{tensor-to-scalar ratio $r=0.1$} primordial gravitational wave background from slow-roll inflation on scales close to the horizon at matter-radiation equality, this is far from true for the new results.

\begin{figure}
	\includegraphics[width=\linewidth]{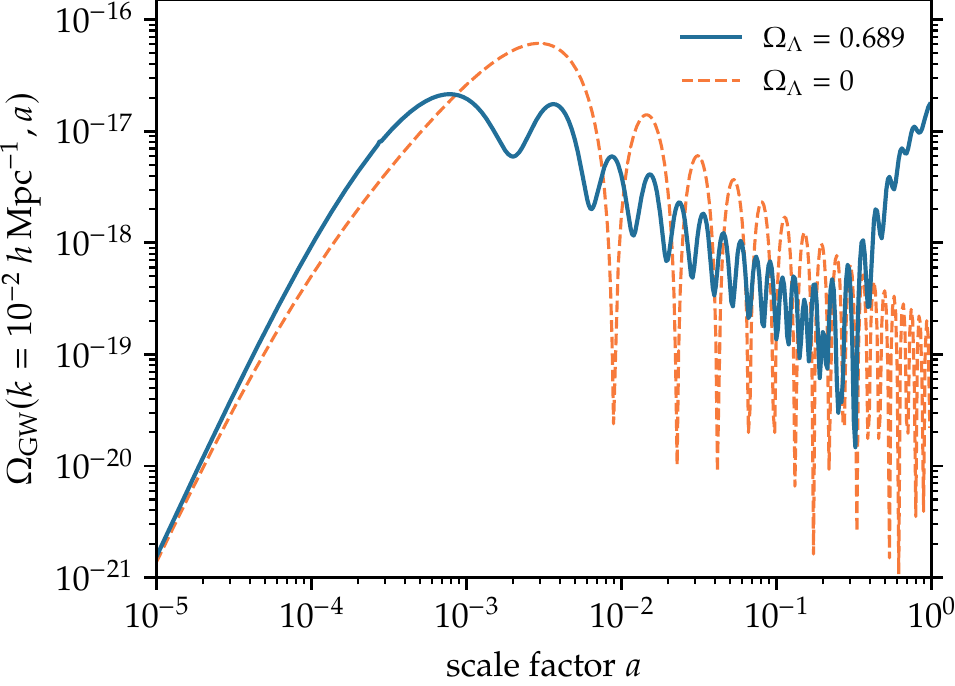}
	\caption{Temporal evolution of $\Omega_\mathrm{GW}(k=10^{-2}\,h/\si{Mpc}, a)$. The mode enters the horizon during matter domination. For both a \textsc{Planck}-like $\Lambda$CDM cosmology (\textcolor{my0}{\bfseries blue}, solid) and an identical one except $\Ol = 0$, $\Om = 1-\Or$ (\textcolor{my1}{\bfseries orange}, dashed), the mode redshifts after horizon entry (${\propto\! a^{-1}}$), as expected during matter domination (cf.~Figure~\ref{fig:Omega_MD_comparison}).
	In presence of a cosmological constant, however, it starts to grow again after $a \approx 2\times10^{-1}$. In addition, the oscillations are suppressed, i.\,e.~there are non-oscillating contributions.}\label{fig:Omega_k1e-2}
\end{figure}

Note, however, that the oscillatory features are still suppressed on large and intermediate scales for the results that do not employ the flawed approximation. This is expected due to non-oscillatory tensor perturbations induced in presence of the cosmological constant. In fact, Figure~\ref{fig:Omega_z0} also includes a plot for $\Omega_\Lambda = 0$, where the aforementioned effect is absent. In Figure~\ref{fig:Omega_k1e-2}, the temporal evolution of a single mode, $k = 10^{-2}\,h\,\si{Mpc^{-1}}$, is shown, both with and without cosmological constant. As expected from~\eqref{eq:Omega_MD_true}, the spectrum redshifts in the latter case, while in the former it starts to rapidly grow again after $a \approx 2\times10^{-1}$. The growth seems to be mainly in a (strongly gauge-dependent) non-oscillatory component, as oscillatory features are increasingly suppressed towards $a=1$.
While we cannot explain this in terms of the analytical de Sitter solution from Section~\ref{sec:de_dom}, we have seen there that non-oscillatory contributions to the tensor perturbations arise in presence of a cosmological constant in conformal Newtonian gauge. In addition, when the cosmological constant becomes the dominant component of the Universe’s energy-momentum content, linear structure starts to dilute again, which might contribute to the sourcing of tensor perturbations. However, for an analytical understanding of the numerically computed late-time growth, the transition between matter and Dark Energy domination needs to be investigated more closely.

\section{Summary and Conclusion}
In this paper, we revisit the late-time scalar-induced gravitational wave background in a $\Lambda$CDM cosmology in light of the breakdown of a commonly applied approximation during cosmological epochs with vanishing speed of sound, $\cssq = 0$. We stress that during matter domination sub-horizon gravitational waves are not free, in contrast to e.\,g.\ radiation domination. While this is well-known, it also implies that one cannot employ the approximation $h_{ij}' \propto k\,h_{ij}$, which is commonly applied implicitly in the definition of $\Ogw$, an issue that was briefly addressed also by \textcite{Ali2021}. We reconsider previous analytical computations for a matter-dominated universe with flat primordial matter power spectrum, showing that the erroneous assumption leads to a strong overestimation of the gravitational wave energy density spectrum.

Furthermore, the presence of a cosmological constant introduces non-oscillating and therefore strongly gauge-dependent contributions to the spectrum. The sources are first-order perturbations and our results are independent of whether they are modeled in relativistic or Eulerian perturbation theory. However, one cannot simply take a fluid energy-momentum tensor and use it as a source for the first-order gravitational wave equation, as this would miss effective source terms arising at second order in relativistic perturbation theory.

In order to account for the transition between matter and Dark Energy domination, we numerically compute the spectrum for a \textsc{Planck}-like cosmology, reconsidering the results of \textcite{Baumann2007} with the aforementioned in mind. While the numerical results generally agree with the analytical findings, we observe a non-oscillating, growing contribution to the spectrum that might be due to the transition between two cosmological epochs that has not been analytically accounted for. This could be a subject of further research.

While our results were obtained in conformal Newtonian gauge, we argue that restricting to purely oscillating contributions to the tensor perturbations would yield approximately gauge-invariant results, along the lines of Refs.~\cite{Matarrese1998, Ali2021, Domenech2021}. However, the fact that the non-oscillating tensor perturbations are strongly gauge-dependent does not mean that they are completely irrelevant. As \textcite{Domenech2021_review} points out, a proper understanding of a second-order detector response would clarify the measurable signal of induced gravitational waves, constituting another important direction for further research.

\begin{acknowledgments}
	This work was supported by the Deutsche Forschungsgemeinschaft (DFG, German Research Foundation) under
	Germany's Excellence Strategy EXC 2181/1 - 390900948 (the Heidelberg STRUCTURES Excellence Cluster).
	For multi-dimensional numerical integrals we used the \emph{Cuhre} algorithm from the CUBA library~\cite{Hahn2005}.
\end{acknowledgments}

\appendix
\section{Relativistic and Eulerian First-Order Scalar Perturbations}\label{app:first_order_pert}
In order to make this paper self-contained, we summarize the relevant results of first-order (scalar) perturbation theory in the following.
Cosmic structure can be treated in terms of small perturbations on top of a homogeneous and isotropic background. First-order perturbation theory of Einstein's field equations of general relativity around a spatially flat FLRW background with perfect fluid energy-momentum results in a set of equations for the gauge-invariant Bardeen potential~\cite{Mukhanov1992},
\begin{align}
	\laplacian\Phi - 3\H(\Phi'+\H\Phi) &= 4\pi G a^2 \delta\rho\label{eq:generalized_poisson}\\
	\partial_i(a\Phi)' &= -4\pi G a^2 (\bar\rho+\bar p)\,\delta {u_\parallel}_i\hspace{4ex}\label{eq:rel_vel}\\
	\Phi'' + 3\H\Phi'+(2\H'+\H^2)\Phi &= 4\pi G a^2\delta p\label{eq:rel_3},
\end{align}
in conformal Newtonian gauge. With a suitable choice of gauge-invariant variables replacing $\delta \rho$, $\delta p$ and $\delta u_\parallel^i$, however, these expressions are valid for arbitrary gauges. $\delta u_\parallel^i$~denotes the scalar (or divergence) part of the velocity perturbation. Since the vorticity quickly decays at first order under the above assumptions, we can safely reduce the first-order velocity perturbation to the scalar part at late times~\cite{Malik2009}.
\eqref{eq:generalized_poisson} and~\eqref{eq:rel_3} can be combined into
\[\label{eq:scalar_evolution}
	\Phi''+3(1+\cssq)\H\Phi'+\qty(2\H'+(1+3\cssq)\H^2-\cssq\laplacian)\Phi = 0,
\]
where entropy perturbations were neglected (i.\,e. adiabatic perturbations were assumed) and $\cssq = \delta p / \delta \rho$.

Sufficiently below the background curvature scale $\H$, the background is approximately flat. In this regime, and since only small perturbations are assumed, the Newtonian limit of gravity is valid for non-relativistic matter. It is therefore common to model linearly evolving late-time structure as fluid obeying the linearized Euler, continuity and Poisson equations. In comoving coordinates, they read~\cite{Peacock1998,Bernardeau2002}
\begin{align}
	a \vec{u}' + 2\,\H a\vec{u} &= -\frac{\nabla \delta p}{\bar\rho}-\nabla\Phi\label{eq:leuler}\\
	\delta' + a\nabla\cdot\vec{u} &= 0\label{eq:lcontinuity}\\
	\laplacian\Phi &= 4\pi G a^2\bar\rho\,\delta\label{eq:poisson}
\end{align}
respectively, where $\delta \equiv \delta\rho/\bar\rho$ and $\vec{u}$ denotes the comoving peculiar velocity. Note that the Poisson equation~\eqref{eq:poisson} is indeed the small scale limit, $ \H^2 \ll k^2$, of~\eqref{eq:generalized_poisson}. From the first two equations it can be concluded that again vorticity decays~\cite{Bernardeau2002}. The above equations can be combined to yield, in the adiabatic case, the growth equation
\[\label{eq:matter_growth}
	D_+'' + \H D_+' = \qty(4\pi Ga^2 \bar\rho+\cssq\nabla^2)\,D_+,
\]
where $D_+$ is the temporal part of the density contrast, $\delta(\vec{x},\eta) = D_+(\eta)\,\delta(\vec{x},\eta_\mathrm{i})$, normalized to unity at some initial time $\eta_\mathrm{i}$ unless stated otherwise. For non-relativistic matter, the Poisson equation implies ${\Phi \propto D_+/a}$. In the pressureless case we have $w=\cssq=0$ and~\eqref{eq:matter_growth} is exactly equivalent to~\eqref{eq:scalar_evolution} upon subtraction of the second Friedmann equation.

\section{Green's Functions to the Gravitational Wave Equation}\label{app:greens}
In Fourier space, the gravitational wave equation~\eqref{eq:second_order_gws} is equivalent to~\cite{Baumann2007}
\[\label{eq:subst_wave}
	H''_{ij} + \qty(k^2- \frac{a''}{a}) H_{ij} = 4a\trtr{\mathcal{S}}_{ij},
\]
where $H_{ij} \equiv a\,h_{ij}$ and the source are understood as functions of $\vec{k}$ and $\eta$. This is a second-order inhomogeneous linear ordinary differential equation. Specifying initial conditions at some time $\eta_\mathrm{i}$, the corresponding initial value problem on the domain $\eta\in[\eta_\mathrm{i},\infty)$ can be solved using the Green's function method, yielding~\eqref{eq:h_conv}. The Green's function $g_k$ is given in terms of two linearly independent solutions $v_k$, $u_k$ to the homogeneous equation to~\eqref{eq:subst_wave},
\[
	g_k(\eta,\tilde\eta) = \frac{u_k(\tilde\eta)v_k(\eta)-u_k(\eta)v_k(\tilde\eta)}{u_k(\tilde\eta)v_k'(\tilde\eta)-u_k'(\tilde{\eta})v_k(\tilde{\eta})}.
\]
The denominator defines the Wronskian $W[u_k,v_k](\tilde\eta)$. For modes well within the horizon, $\H^2 \ll k^2$, the term proportional to $a''/a$ is negligible and the homogeneous equation is simply a harmonic oscillator. A simple set of homogeneous solutions is given by $\sin(k\eta)$ and $\cos(k\eta)$, such that
\[\label{eq:greens_sub_rad}
	g_k(\eta,\tilde\eta) = \frac{\sin(k(\eta-\tilde\eta))}{k}.
\]
Consider a general cosmic fluid with constant equation of state $w \neq -1/3$. From Friedmann's equations we have
\[
	\qty(\frac{a'}{a^2})^2 \propto \rho \propto a^{-3(1+w)} \implies a\propto\eta^{\frac{2}{1+3w}},
\]
and thus the homogeneous part for~\eqref{eq:subst_wave} reads
\[\label{eq:general_homogeneous}
	H_{ij}'' +  \qty(k^2- \frac{2-6w}{(1+3w)^2} \frac{1}{\eta^2}) H_{ij} = 0.
\]
For radiation domination, $w=1/3$, this is again a harmonic oscillator, so~\eqref{eq:greens_sub_rad} is recovered~\cite{Baumann2007}.

Introducing the substitution $H_{ij}(\eta) = \sqrt{\eta}f(\eta)$, multiplying with $\eta^{\frac{3}{2}}$ and using $\partial_\eta = k\partial_{k\eta}$, equation~\eqref{eq:general_homogeneous} takes the form of Bessel's differential equation,
\[
	\eta^2\,\partial^2_\eta f + \eta\,\partial_\eta f + \qty((k\eta)^2- \qty(\frac{3-3w}{2+6w})^2) f = 0.
\]
Two linearly independent solutions are therefore given by Bessel functions of first and second kind, and resubstitution yields~\cite{Domenech2020}
\[\label{eq:general_hom_solution}
	u_k(\eta) = \sqrt{\eta}\,J_{\frac{3-3w}{2+6w}}(k\eta), \quad v_k(\eta) = \sqrt{\eta}\,Y_{\frac{3-3w}{2+6w}}(k\eta).
\]
For both non-relativistic matter, $w=0$, or a cosmological constant, $w=-1$, the homogeneous solutions~\eqref{eq:general_hom_solution} become proportional to~\cite{Baumann2007}
\[
	u_k(\eta) = \eta\,j_{1}(k\eta), \quad v_k(\eta) = \eta\,y_{1}(k\eta),
\]
with the spherical Bessel functions $j_1$ and $y_1$. The Wronskian is $W[u_k,v_k](\tilde\eta) = k^{-1}$~\cite[cf.][]{Abramowitz1964} and the resulting Green's function
\begin{multline}\label{eq:greens_mat_lam}
	g_k(\eta,\tilde\eta) = \qty(1 + \frac{1}{k^2\eta\tilde\eta})\frac{\sin(k(\eta - \tilde\eta))}{k} \\- \frac{(\eta - \tilde\eta)\cos(k(\eta - \tilde\eta))}{k^2\eta\tilde\eta}.
\end{multline}
One can see that the leading order term in $k^{-1}$ is just~\eqref{eq:greens_sub_rad} and thus the small scale limit is recovered as $k\to\infty$. For both~\eqref{eq:greens_sub_rad} and~\eqref{eq:greens_mat_lam}, small scales are suppressed, justifying the assumption that the sourcing of a mode is strongest just around horizon entry.

\bibliography{bibliography.bib}

\end{document}